# Jigsaw Cryptanalysis of Audio Scrambling Systems

Hamzeh Ghasemzadeh[1*], Mehdi Tajik Khass[2], Hamed Mehrara[3]

[1]Faculty of Electrical and Computer Engineering, Islamic Azad University, Damavand Branch, Tehran, Iran

[2]Faculty of Electrical and Computer Engineering, Tabriz University, Tabriz, Iran

[3]Departments of Electrical Engineering, KNT University, Tehran, Iran

*Corresponding Author:
hamzeh_g62@yahoo.com

**Abstract:** Recently it was shown that permutation-only multimedia ciphers can completely be broken in a chosen-plaintext scenario. Apparently, chosen-plaintext scenario models a very resourceful adversary and does not hold in many practical situations. To show that these ciphers are totally broken, we propose a cipher-text only attack on these ciphers. To that end, we investigate speech permutation-only ciphers and show that inherent redundancies of speech signal can pave the path for a successful cipher-text only attack. For this task different concepts and techniques are merged together. First, Short Time Fourier Transform (STFT) is employed to extract regularities of audio signal in both time and frequency. Then, it is shown that cipher-texts can be considered as a set of scrambled puzzles. Then different techniques such as estimation, image processing, branch and bound, and graph theory are fused together to create and solve these puzzles. After extracting the keys from the solved puzzles, they are applied on the scrambled signal. Conducted tests show that the proposed method achieves objective and subjective intelligibility of 87.8% and 92.9%. These scores are 50.9% and 34.6% higher than scores of previous method.

## 1. Introduction

Recent advances in the field of communications have considerably increased usage of multimedia signals over the wireless channels. Wireless communication channels are prone to different kinds of attacks and therefore their contents should be protected. To address these requirements, some new security services have been proposed. They include watermarking, steganography, and multimedia encryption systems. Each of these services fulfils a different task. Watermarking [1] serves the purpose of copyright protection, steganography implements a subliminal channel [2] and steganalysis tries to detect it [3, 4], and multimedia encryption systems provide confidentiality [5]. Demand for high speed and real time services, large amount of data, constraints on energy consumption, and unique characteristics of multimedia signals have turned multimedia encryption systems into an interesting subject. Reviewing existing literature shows that many papers have been published on design and analysis of these systems [6-8], especially in the image context.

Systems that provide confidentiality for audio channels fall into different categories. In digital encryption systems, the signal is compressed to achieve the desirable bit rate then it is enciphered with an encryption technique. GSM network is an example of this category. On the other hand, scrambling systems take a different path and work directly on the multimedia samples or their transformed counterparts. Scrambling systems have some features which make them more suitable for certain purposes. These features include high speed, low power consumption, excellent voice recognition, independency of the quality of recovered signal from the language and speaker, possibility of directly coupling them to any handset, removing the need for speech compression and modem, and the most important of all, compatibility with the existing analogue and narrowband communication channels [9]. Reviewing the literature shows that there are different methods for realizing this task. While the conventional scramblers manipulate audio signal in the frequency and/or time domain, modern methods rely on transformations. The hopping-window time domain scrambler permutes segments of audio signal in the time domain. On the other hand, the band splitting scrambler, breaks the spectrum of the signal into several sub-bands and then performs permutation and frequency reversion on them [10]. In another work, chaotic system was used to interleave voice packets into different frames of network for real time voice over IP (VoIP) applications [11]. Owing to the dawn of high-speed signal processing hardware, more complex methods have been proposed. This class of scramblers works in transform domains. First, they apply some transformation on the digitized signal. Then, after permuting the resultant coefficients, the inverse transform is applied to create the scrambled signal. Hadamard matrices were used for speech scrambling in [12]. Sakurai et al. [13] proposed a method based on re-arrangement of the coefficient of Fast Fourier Transform (FFT). To improve security of the system, adaptive dummy spectrum insertion technique was used [14]. A method based on discrete cosine transform (DCT) was suggested in [15]. Goldburg et al. investigated the potency of audio scrambling in transform domain [16]. They considered discrete Fourier transform, DCT, Walsh-Hadamard Transform, and discrete Prolate spheroidal transform. Later, It was shown that DCT is the best transform to be used for transform-based encryption [17]. A practical system based on ITU-T G.723.1 speech codec and DCT permutation was proposed in [18]. To increase the effective number of permutations, an scrambling algorithm based on the wavelet packets was proposed [19]. Another method based on the parallel structure of two different types of wavelet with the same decomposition level was proposed in [20]. Tseng et al. argued that most scrambling methods preserve the signal energy, talk spurts, and the original intonation. To solve these problems, they proposed a method based on orthogonal frequency division multiplexing [21]. Li et al. increased the dimension of sample points coordinates and used arbitrary matrices to scramble audio signals [22]. Another work used intractability of the problem of underdetermined blind source separation for speech encryption [23]. Idea of progressive scrambling/descrambling in the wavelet domain and MP3 files was proposed in [24]. This idea allows a person to retrieve different qualities of audio signal, based on the amount of key that he knows. Different approaches for joint multimedia compression and encryption were discussed in [25]. Zeng et al. used compressed sensing ideas to construct a robust scrambling method against active attacks [26]. Linear feedback shift register (LFSR) were used for selective encryption of compressed audio signals [27]. Hierarchical selective encryption of G.729 standard was investigated in [28]. According to this method bit stream was partitioned into the most sensitive and the least sensitive parts. Then, different chaotic maps were used for encryption of each part. Finally, joint scrambling and watermarking of MP3 were discussed in [29, 30].

While different audio encryption and scrambling methods have been proposed, very few works have been published on their security evaluation. First, Goldburg et al. employed a frequency domain vector codebook to estimate a model for space of clear speeches [17]. Then, this model was used to implement a ciphertext-only attack on fixed and varying permutation frequency domain scramblers. Later, this method was extended to cryptanalysis of DFT-based systems [31]. Spectral distance measures between the end of each segment and the start of all other segments were employed for cryptanalysis of time scrambling systems in [32]. Finally, Zhao et al. proposed a technique for solving rectangular jigsaw puzzles and mentioned audio cryptanalysis as a possible application, but the paper neither provided the means for converting the speech into jigsaw puzzles, nor presented any results of such cryptanalysis system [33].

Recently, security of permutation based multimedia encryption systems were studied [7]. The work showed that "most (if not all) permutation-only multimedia ciphers" can completely be broken in a chosen-plaintext scenario. In a chosen-plaintext scenario, the attacker can obtain the ciphertext of any plaintext he wishes [34]. Apparently this model of attack is very powerful and does not hold in many practical situations. In fact, this model is mostly applicable to the public key cryptography systems [34]. Another model of attack is ciphertext-only attack. In this model, the attacker has only access to a set of ciphertexts. Not only this model is more realistic and applicable to any cryptosystem, but also this vulnerability leads to vulnerability to chosen-plaintext model whereas the reverse may not hold. In order to show that permutation based systems are insecure in the practical setting, we presents a ciphertext-only attack on these systems. We show that time domain permutation based audio encryption systems can be broken very efficiently. Also, the proposed attack is independent from key generation mechanism; therefore, no better pseudorandom permutation mapping can be realized to offer a higher level of security against the proposed attack. Through different simulations we show that the proposed method results in acceptable intelligibility even if the scrambling system uses small segment lengths and high frame sizes. Also, the proposed system retrieves quite intelligible speeches even in low values of signal to noise ratios (SNRs).

The main contributions of our work are:
- Continuing on our seminal work [35], we present a practical ciphertext-only attack on time domain audio scrambling system. Therefore, these systems can be considered totally broken. The attack is based on converting scrambled audio signals into two dimensional image puzzles. The suitable transformation and its optimum parameters are also presented.
- We show that the proposed transform has a windowing effect which causes a discontinuity between correct segments. This phenomenon may drastically reduce performance of the system. To solve this issue, we present a novel solution where the boarder samples are estimated. We show that this technique solves the aforementioned problem.
- Both objective and subjective tests are conducted to quantify efficacy of the proposed system. Also, effect of different parameters of the scrambling system, including its segment length, its key size, and SNR of the channel on the intelligibility of descrambled speeches are investigated.

The rest of this paper is organized as follows: Section 2 presents the concepts of hopping window scramblers and STFT. Section 3 elaborates on converting scrambled signal into jigsaw puzzles and then on solving the jigsaw puzzles. Section 4 presents results of simulations and tests. Discussion follows in the section 5 and finally paper concludes in section 6.

## 2. Preliminaries

### 2.1. The hopping window time-domain scrambler

These systems maintain confidentiality through permutation of signal in the time domain. To that end, first, audio signal is split into frames with predefined duration. Then, each frame is further split into $N$ non-overlapping segments, where $N$ is called frame size. Afterwards, segments in each frame are re-arranged according to a permutation key [10]. In order to improve the security, this key should vary for each frame [31].

### 2.2. Spectrogram

STFT is a 2D transform from time domain to frequency–time domain. It visualizes any time varying signal versus its spectrum of frequencies. Principally, the signal is multiplied by a sliding window (typically Hamming). This window sweeps the signal in time domain (usually with some overlap) and then magnitude of the frequency spectrum is calculated using FFT. Considering signal $x(n)$, its STFT representation is defined as [36]:

$$X(m, \omega) = \sum_{n=-\infty}^{+\infty} x[n]. W_m(n). e^{-jwn} \qquad (1)$$

where $W_m(n)$ is a sliding window with non-zero values in the interval of [$m$, $m+\delta$-1] and overlap of $\gamma$ samples. By changing the window size ($\delta$), it is possible to trade off time resolution for frequency resolution. If we use an N-point FFT for calculation of equation (1) and $x(n)$ has $L$ samples, the size of $X(m,\omega)$ can be calculated as [37]:

$$1 \leq k \leq \frac{N}{2}, \qquad 1 \leq m \leq \left\lfloor \frac{L-\gamma}{\delta-\gamma} \right\rfloor \qquad (2)$$

### 3. Proposed Cryptanalysis Method

Speech signal can be predicted both accurately and efficiently [38]. Speech coding techniques, exploit this features to achieve high compression rates. Investigating different clear speeches shows that these characteristics are reflected very clearly in their spectrograms. In this manner, the spectrogram of a clear voice is smooth in both directions (time and frequency) and it exhibits no abrupt transitions. On the other hand, the spectrogram of a scrambled voice shows abrupt transitions in the border of segments and sub-bands. Figure 1 demonstrates these phenomena.

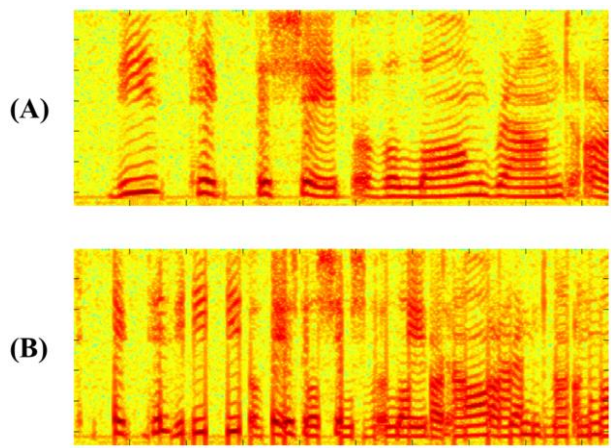

Figure 1.    Comparing continuity of STFT of clear and scrambled speech in the time domain    (A) Clear speech    (B) Scrambled speech

Regarding figure 1, it is possible to distinguish between clear and scrambled signals. This observation can be exploited for cryptanalysis of scrambled voices. In this manner, cryptanalysis of the scrambled signal can be considered as re-arranging its time segments in a fashion that the resulting signal produces a regular and smooth spectrogram. This problem is equivalent to solving the jigsaw puzzle of the scrambled signal. Cutting spectrogram into its pieces is the prerequisite of this step.

### 3.1. Transforming scrambled signal into jigsaw puzzles

According to equation (1), a sliding window is applied on the signal to produce its spectrogram. Although this windowing technique adds time resolution to the FFT, but it makes the transition area between two adjacent segments blurry. Figure 2.A shows this fact more clearly. Let $x(n)$ and $X(m,\omega)$ denote a scrambled signal and its spectrogram, respectively. Extracting pieces of the puzzle from $X(m,\omega)$ is highly problematic. First, exact border of the segments are not clear in the $X(m,\omega)$. Second, since the segments are derived from $X(m,\omega)$, it is very likely that solution of the puzzle solving algorithm be the same as $X(m,\omega)$. To discuss these phenomena more clearly, assume STFT is implemented using a sliding window with duration of $\delta$ samples and the maximum value of overlap ($\gamma=\delta$-1). Furthermore,

$x(n) = \{x_{i,j}\}, 1 \leq i \leq T, 1 \leq j \leq L$

n = j + (i − 1) × L     (3)

where *i*, *j*, *L*, and *T* denote segment index, sample index, segment length, and total number of segments, respectively. While the hamming window is sliding from segment k to *k+1*, some transition steps occur. In these steps, head of the sliding window covers the initial samples of *k+1*th segment whereas its tail is on the remaining samples from *k*th segment. These transition steps are:

Lk − δ + 1 ≤ m ≤ Lk − 1 , 1 ≤ k ≤ N − 1     (4)

According to equation (4), the width of this transition region is equal to *δ-1* pixels. Furthermore, as m is moving from *Lk-δ-1* to *Lk-1*, gradually impact of samples from *k*th segment on the $X(m,\omega)$ decreases and impact of samples from *k+1*th segment increases. Consequently, instead of occurring in a single pixel which results a sharp and clear border of each segment, the transition area from the *k*th segment to *k+1*th segment would be a smooth and blurred region. Therefore, the exact borders of segments are not clear. This phenomenon is highlighted in the figure 2.A.

Reviewing literature on solving jigsaw puzzles shows that border pixels play a vital role in the solution of puzzle. Among all possible arrangements, these algorithms choose the arrangement with smoothest overall transition in the boarders. As we argued, STFT smooths down the spectrogram of the encrypted signal. Since the criteria for puzzle solving algorithm is also the smoothness, STFT may mislead the algorithm and the solved puzzle may be the same as the spectrogram of encrypted speech.

To solve these problems, first, we cut the scrambled signal into distinct segments. Then STFT of each segment is calculated separately. In this fashion both of aforementioned problems are solved. Figure 2 compares segmented spectrogram and conventional spectrogram of a portion of scrambled speech.

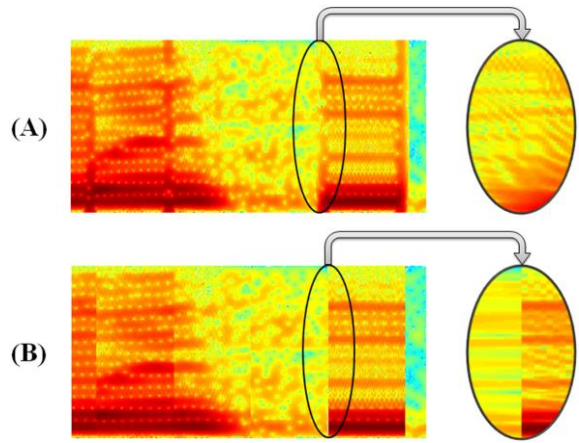

Figure 2.     Cutting spectrogram of speech into pieces of jigsaw puzzle     (A) Conventional spectrogram (B) Segmented spectrogram

Segmented spectrogram can solve the above-mentioned problems, but it has its own drawback. Let $x_{k,1}$, $x_{k,2}$, …, $x_{k,L}$ denote samples in the *k*th segment. Investigating mathematical representation of the segmented spectrogram shows that middle samples and border samples act differently. To be more specific, if γ=δ-1, then middle samples contribute in δ sliding windows ($W_m(n)$). That is, if a middle sample is changed then values of δ columns of segmented spectrogram would change. But in the border samples, this effectiveness gradually reduces to one column. Let *j* denotes index of a sample in the *k*th segment, equation

(5) shows value of effectiveness in the segmented spectrogram. In other words, for *j*th sample, Φ(j) shows the number of windows which it can affect.

$$\Phi(j) = \begin{cases} j & j < \delta \\ \delta & \delta \leq j \leq L - \delta \\ L - j + 1 & j > L - \delta \end{cases} \quad (5)$$

Different values of effectiveness in the border samples cause a discontinuity between consecutive segments even if they are re-arranged correctly. This effect is illustrated in the figure 3.A. In the next section a novel method is proposed to mitigate this problem.

### 3.2. Estimating border samples

Let $E(x, \pm l)$ denotes an ideal algorithm that can accurately predict both l-past and l-future samples of the sequence *x*. Now we calculate:

$$y = E(x, \pm \delta - 1) = [x'_{-\delta+1}, \dots, x'_{-1}, x_1, \dots, x_L, x'_{L+1}, \dots, x'_{L+\delta-1}] \quad (6)$$

which is a sequence with L+2δ-2 samples. Effectiveness values of the middle indices ($1 \leq j \leq L$) of this new sequence are:

$$\Phi(j) = \delta \quad 1 \leq j \leq L \quad (7)$$

In other words, in the segmented spectrogram of sequence *y*, all samples of sequence *x* will have the same value of effectiveness. A simulation is conducted to justify that ideal estimation removes discontinuity between segmented spectrogram of the sequences *x*. Figure 3 compares the segmented spectrogram of correct arrangement of pieces with ideal estimation and without it.

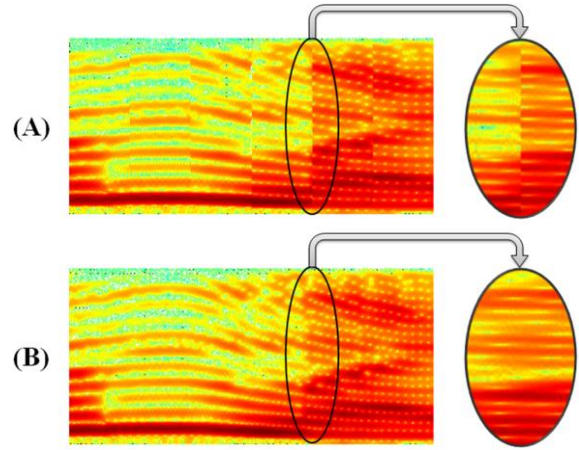

Figure 3.     Removing discrepancies between correct neighbors in the segmented spectrogram (A) Without estimation     (B) With ideal estimation

### 3.3. Recursive least squares estimation

Let *x(n)* and *d(n)* denote input and output samples of a system. Recursive least squares estimation (RLS) is an adaptive transversal filter of the form [39]:

$$\hat{d}(n) = \sum_{k=0}^{M} w_n(k) x(n - k) \quad (8)$$

such that $x_n = [x(n), x(n-1), \dots, x(n-M)]$ is the vector containing the *M+1* most recent samples of input signal and *M* is the order of filter. Design objective is estimating weights of the filter at each time *n* ($w_n$), in a way that the cost function is minimized. The cost function is defined as [39]:

$$C(w_n) = \sum_{i=0}^{n} \lambda^{n-i} \cdot e(i)^2 \quad (9)$$

$\lambda$ is called the forgetting factor of the algorithm and it reduces the influence of old data. Furthermore, the error signal (*e*) is calculated as [39]:

$$e(n) = d(n) - \hat{d}(n) \quad (10)$$

### 3.4. The Proposed method

According to Kerckhoffs's principle [40], it is assumed that security of the system relies only on its key. Therefore, all the parameters of the scrambling system including its segment length and frame size are known a priori. Block diagram of the proposed method is presented in figure 4.

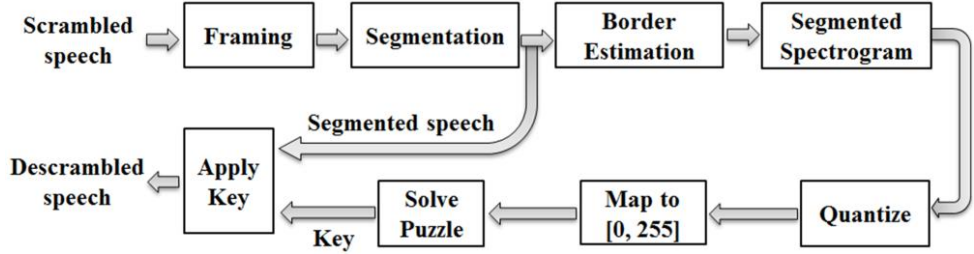

Figure 4.  Block diagram of the proposed cryptanalysis method

First, according to the frame length, audio signal is split into non-overlapping frames. Then, each frame is further divided into non-overlapping segments. After that, both (δ-1)-past and (δ-1)-future samples of the border samples in each segment are estimated, where δ is the length of window in the STFT transform. Then, segmented spectrogram is applied on each estimated segment to transform them to time-frequency domain. At this stage, each segment has transformed into a two dimensional signal. The resulting signal is quantized and then all segments within the same frame are jointly mapped onto values between zero and 255 (corresponding to 256 grey scale levels). At this point, a scrambled jigsaw is obtained for each frame of the scrambled speech. These puzzles are fed into puzzle-solving algorithm, where a combination of different image processing techniques and search methods are employed to find the best solution of the puzzle. Finally, the key of each frame is extracted from the solved puzzle and it is used to descramble the encrypted frame.

### 3.5. Solving the Puzzle

Puzzle solving is the art of pattern recognition and combinatorial optimization. There are many different practical applications that can be reduced to jigsaw puzzles [41]. Rectangular jigsaw puzzles are a class of puzzles where the borders of all pieces have rectangular shape. Solving rectangular jigsaw puzzles can be broken into two main tasks. First, a proper metric should be devised to assess how well two pieces match each other. This problem can be tackled by using signal processing techniques. Result of this analysis can be represented as a distance value. Second, the space of all possible arrangements should be searched efficiently. This problem is a combinatorial problem and leads to an optimization task.

#### 3.5.1. Distance Function:

To find the best arrangement of puzzle pieces, it is necessary to formulate a measure of correctness. This task can be accomplished by comparing the shape, colour, texture etc. of border pixels. Let $I_i(x,y)$ be the spectrogram image of an audio segment with size of $m \times n$. A suitable distance function between $I_1$ and $I_2$ is the root mean square (RMS) value of differences between their border pixels. In the previous section we showed that an ideal estimation removes discontinuity of border in the segmented spectrogram of true neighbours. Unfortunately, such ideal algorithm does not exist in practice; thus, even for true neighbours, a small discrepancy between the borders would remain. To mitigate this problem, we let image $I_1$ to slide along image $I_2$ vertically for small number of pixels ($\beta \in \{0,1,\dots,7\}$). Also, we let two pieces penetrate into each other for small number of pixels ($\alpha \in \{0,1,2,3\}$). This concept is illustrated in figure 5. We use the lowest distance as an estimation of true distance between two pieces $I_1$ and $I_2$.

$$D = \min_{\beta,\alpha}\{\sqrt{\sum_{y=1}^{y=n-\beta}(I_1(m-\alpha,y+\beta)-I_2(\alpha,y))^2}\} \quad (11)$$

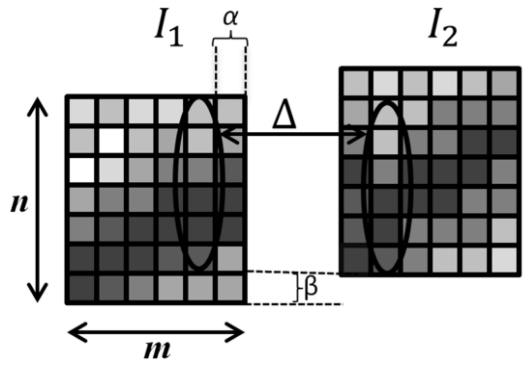

Figure 5. For caculating distance between two images, they are slid horizontaly and vertically untill the best match is found.

Also, equation (11) and figure 5 consider only when I2 is slid upward along I1 piece. The actual distance was calculated as the minimum value of both upward and downward sliding. Figure 6 represents a better insight about the proposed distance.

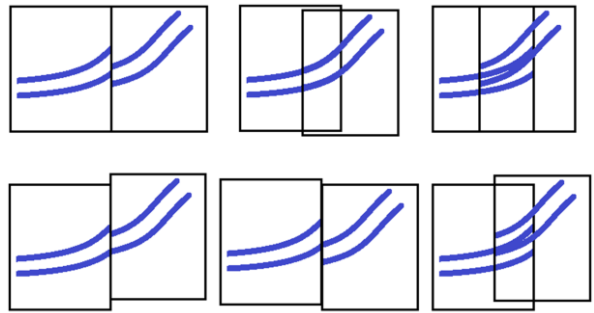

Figure 6. Two pieces of the puzzle are slid along each other, and penetrated until the best match is found

### 3.5.2. Solving puzzle algorithm:

Distance function of the previous section was employed to calculate net cost for every possible arrangement of pieces. This cost is equal to:

$$\text{Dist} = \sum_{i=1}^{N-1} D(I_i, I_{i+1}) \quad (12)$$

Now, solving the jigsaw can be expressed as an optimization problem [42]:

$$S = \min\{\text{Dist}\}_\Omega \quad (13)$$

where, *N* is the frame size and Ω is the space of all possible arrangements of all pieces.

Let *k* and *L* denote duration of scrambled speech (in minutes) and segment length, respectively. If the scrambling key is changed for each frame, equation (14) calculates cardinality of the search space.

$$|\Omega| = \frac{k \times 60}{L \times N} \times (N!) \tag{14}$$

To put the size of key space (|**Ω**|) into a better perspective, this value was calculated for *k*=5 minutes, *L*=40ms, and different frame sizes (*N)*. The result is presented in table 1.

**Table 1 size of key space**

| N | |Ω|(bit) |
|---|---|
| 6 | 20 |
| 8 | 26 |
| 10 | 32 |
| 12 | 39 |
| 14 | 46 |

According to table 1 the size of key space is very huge; thus, finding the accurate answer in an efficient manner is very crucial for this method to work. Reviewing literature shows that this task can be accomplished in different ways. For example, different metahuristic algorithms were employed to solve this problem [42]. Another method to solve this problem used minimum spanning tree[43]. Unfortunately these methods may get trapped in local minima of the equation (13). Recently, a method based on branch and bound (B&B) technique was proposed [41]. It was shown that this algorithm finds the global minima of the equation (13) with acceptable search complexity. This algorithm employs a tree for systematic and efficient search of key space. At each step, it selects one node and expands the tree from it. Furthermore, it uses both upper and lower bounds to prune the tree in the most effective way. The best solution found so far (Opt) and the minimum weight arborescence of directed graph [44] were used as the upper and the lower bounds. A simplified schematic of this method is presented in figure 7.

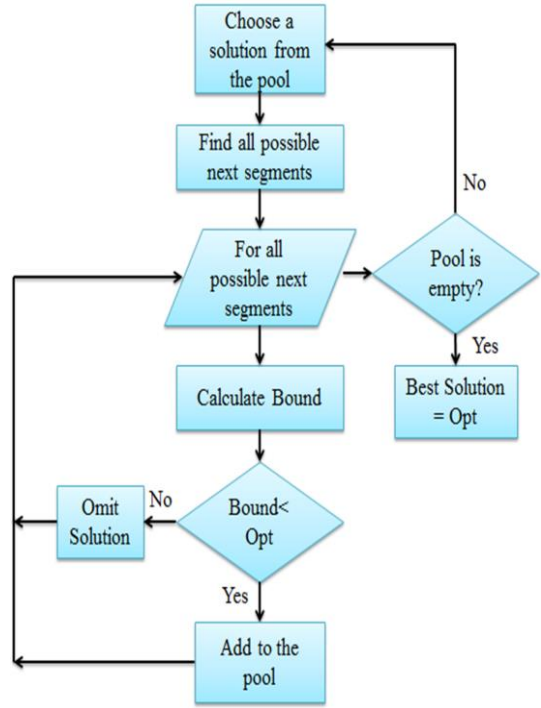

Figure7.   Solving jigsaw puzzle based on B&B method [41]

## 4. EXPERIMENTAL RESULTS

This section investigates efficacy of the proposed method. To that end, time domain hopping window scrambler was simulated using MATLAB. Table 2 presents parameters of the simulated system.

Table 2. Parameters of scrambling system

| Time domain hopping window | |
|---|---|
| Parameter | value |
| Frame Size (N) | 8 |
| Segment Length (L) | 40ms |
| Fs | 8Khz |

### 4.1. Performance criteria:

The core of the proposed method is solving the resulted puzzle. Any error in the solution of the puzzle is directly reflected into the descrambled speech. Our main objective is to achieve the highest value of intelligibility in the descrambled signal. Intelligibility is a subjective measure; therefore, the objective accuracy criteria proposed in [41] was employed. This metric progressively divides the found solution into different sub-blocks and compares them with sub-blocks of correct solution. In this manner, concurring between larger sub-blocks of found solution and correct solution will result in higher score. Let us denote sub-block that starts from *i*th position in the correct solution and found solution with $B_{Ci}$ and $B_{Fi}$, respectively. Now, define $S_n$ as the total number of $B_{Ci}$ and $B_{Fi}$ that are the same:

$$S_n = \sum_{\forall i} \sum_{\forall j} \delta(B_{Fi} = B_{Cj}) \qquad (15)$$

where δ is Dirac delta function. Finally, accuracy was calculated as the weighted sum of $S_n$ for all possible values of *n*.

$$Ac = \frac{\sum_{n=1}^{N} S_n \times n}{\sum_{n=1}^{N}(N - n + 1) \times n} \quad (16)$$

### 4.2. Tests Methodology:

The tests were conducted on two different databases. For the first database 1000 samples were selected randomly from TIMIT [45] database. These samples were divided into two sets of 400 and 600 excerpts for parameter optimization and conducting objective evaluation of intelligibility. For the second database, we used seven Persian speech samples with total duration of 90 seconds. They included both male and female speakers and they were recorded in soundproof room with sampling frequency of 44100 and bit resolution of 16. The second database was used for conduction subjective evaluation of intelligibility. It is noteworthy that, both databases were subjected to the following processes. All excerpts were down-sampled from their original frequency to 8000 Hz. Also, silent portions of the speech have no intelligibility; therefore, a simple energy based voice activity detection method (VAD) was employed to detect and remove silent portions from all samples.

Scrambled samples were produced by feeding each signal into simulated scrambling system. Also, the permutation key was changed for each frame. Finally, the produced cipher voices were descrambled with different methods.

### 4.3. Optimizing parameters:

There are some parameters associated with the proposed method that need to be optimized. STFT has two important parameters: window size and length of overlap. Also, filter order and forgetting factor in the RLS estimation should be determined. An exhaustive search over these four parameters was conducted on the first set of files from TIMIT to achieve the highest value of accuracy. The optimized parameters are shown in table 3. According to table 3 the best results are achieved when STFT is used with window size of 60 and high amount of overlap. Also, the estimation system uses 52 samples for estimating a new sample and it uses high value for the forgetting factor.

Table 3. The Optimum value of parameters

|      | Parameter         | value |
|------|-------------------|-------|
| STFT | Win Size          | 60    |
|      | Overlap Value     | 51    |
| RLS  | Order             | 52    |
|      | Forgetting factor | 0.97  |

### 4.4. Intelligibility of descrambled speech:

To measure efficacy of the proposed system, intelligibility of descrambled samples is measured both objectively and subjectively. In the objective tests, the second set from TIMIT database was used. We used equation (16) to measure objective intelligibility of descrambled speeches. Table 4 compares average intelligibility of different methods.

Table 4. Objective Intelligibility of descrambled speeches

| Method       | Accuracy (mean±std) | Reference |
|--------------|---------------------|-----------|
| **Puzzle + RLS** | 87.8±23.0%          | This work |
| **Puzzle**   | 73.9±31.8%          | [35]      |
| **FWLSD**    | 36.9±37.3%          | [31]      |

According to table 4, the proposed method achieves score of 87.8% while this value for puzzle without estimation and FWLSD methods are 73.9% and 36.9% respectively.

For measuring actual values of intelligibility a set of subjective test was conducted. To that end, all seven Persian samples were descrambled by basic puzzle method [35], system of figure 4, and Frequency Weighted Log Spectral Distance (FWLSD) method proposed in [32]. This resulted in a total of 21 different samples. We divided these files into tree distinct sets, such that each set contained three excerpts descrambled with one method and two samples from other two methods. Then, each set was given to five different persons and they were asked to listen to all samples and transcribe those words that they can understand. With the help of data gathered from these fifteen people (6 female and 9 male), subjective intelligibility was calculated. First, gathered data was divided into three categories according to their method of descrambling. Subjective intelligibility of each method was calculated as the average value of correct transcription of words over each category. Table 5 presents the results.

**Table 5. Subjective Intelligibility of descrambled speeches**

| Method | Accuracy (mean±std) | Reference |
|---|---|---|
| **Puzzle + RLS** | 92.9± 4.5% | This work |
| **Puzzle** | 83.5±10.3% | [35] |
| **FWLSD** | 58.3±14.2% | [32] |

According to table 5 on average, 92.9% of the words descrambled with the proposed method were intelligible. On the other hand, only 83.5% and 58.3% of the words descrambled with puzzle without estimation and FWLSD methods were intelligible.

### 4.5. System parameters and intelligibility of descrambled speech:

To measure effect of scrambler parameters on different descrambling methods, a set of tests were carried out. These tests can demonstrate limitations and potency of each method. In the first test, frame size (N) of scrambler was varied and then scrambled samples were descrambled with different methods. Finally, objective intelligibility of descrambled samples was measured. In the second test, a scrambler with 8 segments per frame with different segment lengths (L) was simulated. After descrambling samples with different methods, their objective intelligibility was measured. Figure 8 shows results of these simulations.

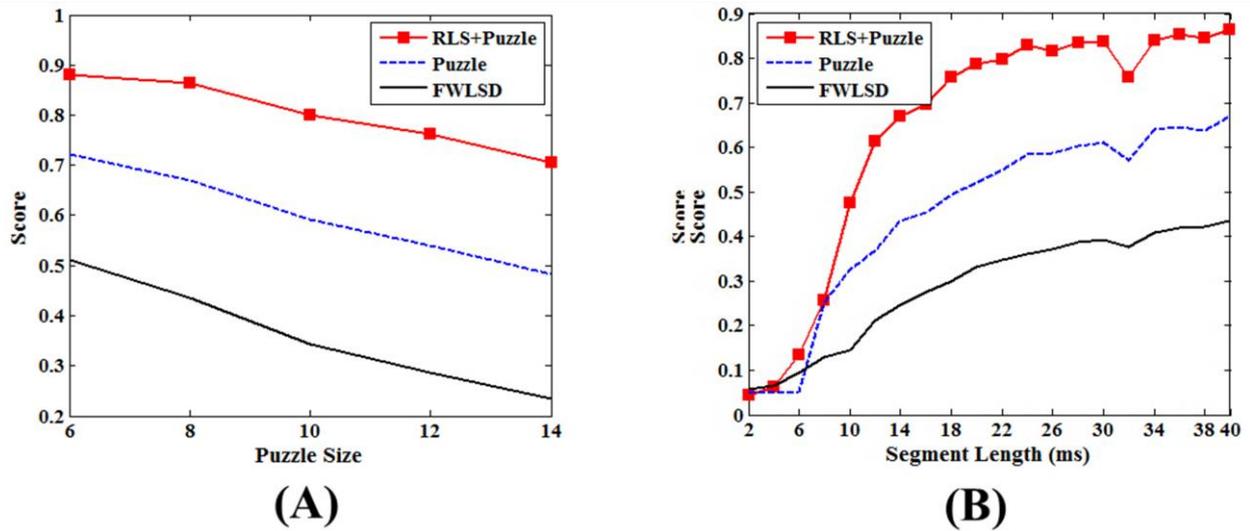

Figure8. Effect of scrambler parametrs on the inteligibilty of descrambled speech.

A: Effect of frame size   B: Effect of segment length

According to figure 8.A when the frame size increases from 6 to 16, intelligibility of RLS+puzzle, Puzzle, and FWLSD methods drops to 0.7, 0.48, and 0.23. Also, intelligibility of FWLSD method decreases faster than RLS+puzzle method. These observations justify that estimation improves performance of the system considerably. Referring to figure 8.B, we observe that the proposed method (RLS+puzzle) with segment length of 10ms can provide better intelligibility than FWLSD method with segment length of 40ms. The same comparison can be made between RLS+puzzle and puzzle methods.

Finally, let 0.5 be the minimum level of acceptable intelligibility. Figure 8.A shows that FWLSD method is acceptable for very small frame sizes (L<6). On the other hand, the proposed method remains in the acceptable range even for big frame sizes (L>14). Also, figure 8.B shows that FWLSD needs large amount of data (N>40ms) to achieve acceptable level of intelligibility. On the other hand, RLS+puzzle needs very small amount data (N = 10ms) to achieve acceptable level of intelligibility. Based on these observations we can conclude that the proposed method (RLS+puzzle) has fewer limitations than other methods.

### 4.6. Effect of noise on intelligibility of descrambled speech:

In practical situations different types of noise are added to the voice. These noises may be categorized into two different types. Source noise mixes with the voice before its scrambling. This noise represents ambient noise and noise of the scrambler system. On the other hand, channel noise mixes with the scrambled voice as it is transmitted through the communication channel. To measure effect of both noises, a set of tests were carried out. In this regard, noise was modelled as an additive white Gaussian noise. Figure 9 shows objective intelligibility of different methods for different power of noises.

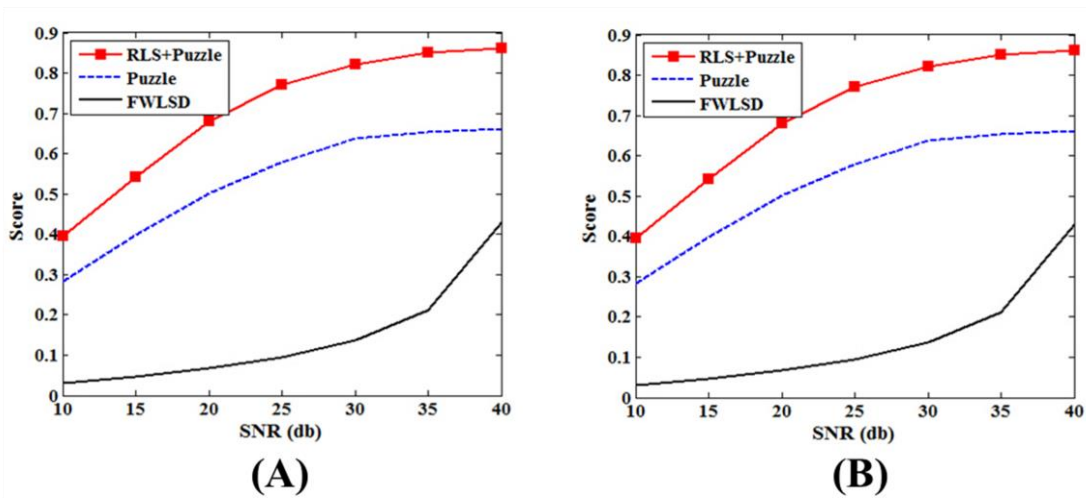

Figure9 Effect of noise on the inteligibilty of descrambled speech.

A: Effect of source noise   B: Effect of channel noise equal. We know that white noise is a stationary random process. Therefore, its statistics remains constant over the time and hence both noises would have the same effect on the intelligibility of descrambled speeches. Comparing results of the different methods demonstrates that intelligibility of the proposed method (RLS+puzzle) with signal to noise ratio (SNR) of 10db is almost equal to the FWLSD method with SNR of 40db. Also, if 0.5 is the minimum level of acceptable intelligibility, then Figure 8.A shows that intelligibility of FWLSD method is acceptable only at very high SNRs (**SNR ≥ 40ms**). On the other hand, the proposed method remains in the acceptable range even for low values of SNRs (**SNR ≥ 15db**). Based on these observations we can conclude that the proposed method has fewer limitations than other methods.

## 5. Discussion

Audio signal has high amount of redundancies. These redundancies could be exploited for cryptanalysis of scrambling systems. According to figure 1 these redundancies are reflected in the spectrogram very well. Comparing spectrogram of clear and ciphered samples shows that, there is no abrupt transition in the spectrogram of clear samples and they are smooth images in both dimensions. Based on these observations, cryptanalysis of audio scrambling system was formulated as a puzzle solving problem. The first step of this method was creating the jigsaw puzzles from scrambled samples. To that end, it was argued that STFT should be applied on each segment separately. Later, it was shown that if segmented spectrogram is used, some discrepancies between borders of true pieces exist. These discrepancies were the result of different values of effectiveness in the border samples. We incorporated concept of estimation in the system to ameliorate these discrepancies. Furthermore, to mitigate effect of non-ideal estimation, distance function was modified to further alleviate effect of any remaining discrepancies. In this regard, the selected pieces were slid in front of each other to find their best match.

Our experiments showed that the proposed method had better performance. According to table 4 in the objective test, the proposed method achieves intelligibility of 87.8% while this score is 73.9% and 36.9% for puzzle without estimation and FWLSD methods. Furthermore, intelligibility of the proposed method had lower variances; thus, not only it results in more intelligible samples, but also its intelligibility does not deviate very much. We believe that higher values of intelligibility of the puzzle-based methods were direct consequence of STFT transform. Investigating formula of FWLSD [46] shows that, it only extracts frequency properties of different sub-bands of audio signal. On the other hand, spectrogram transform reveals both time domain and frequency domain redundancies of the signal. In other words, proposed method exploited redundancies of the speech signal more efficiently.

Comparing tables 4 and 5 shows higher value of intelligibility in the subjective tests. Conducted objective tests had two important properties. First, it considered one frame at a time. Second, only orders of the segments were investigated. However, in the subjective tests both ear and brain plays role. For example, post-masking and pre-masking properties of human auditory system [47] may mask some of the erroneous patterns. Furthermore, brain learns structure of the language and can easily predict and guess portions of sentences in everyday life. Thus, it is very likely that brains of test subjects have used their ability to fill some of the gaps and to find the most suitable words. Therefore, higher value of subjective intelligibility is justified.

## 6. Conclusion

This paper addressed the absence of a through security analysis of scrambling systems. To that end, security of hopping window time domain scrambler was investigated. It was shown that high redundancies of audio signals were reflected in their spectrograms very efficiently. Based on this transformation, cryptanalysis problem was transformed into solving rectangular jigsaw puzzles. Furthermore, integrating estimation technique into the system improved results of the system. Finally, the proposed method was compared with FWLSD which to the best of our knowledge is the only existing cryptanalysis method for these systems. Both subjective and objective tests were carried out to compare intelligibility of different methods. Objective and subjective tests showed that the proposed method achieved intelligibility scores of 87.8% and 92.9%, respectively. These scores were 50.9% and 34.6% higher than scores achieved by FWLSD method.

If 0.5 is defined as the minimum level of acceptable intelligibility, then according to results of figures 8 and 9 the proposed method has the following limitations. In the proposed method, segment length should be larger than 10ms and SNR of scrambled speeches should be at least 15 db. It is noteworthy that these parameters should be higher than 40ms and 40db in the FWLSD method. As the future work we are working to extend this method to other scrambling techniques.

**Acknowledgments**
The authors would like to thank all the participants in the subjective tests.